\documentclass[preprint,aps,prd,superscriptaddress,showpacs,nofootinbib,noshowkeys,floatfix,preprintnumbers]{revtex4}
\usepackage{times}

\usepackage[utf8]{inputenc}

\usepackage{graphics,graphicx}
\usepackage{amsmath, amssymb}
\usepackage{multirow}
\usepackage{longtable}
\usepackage{color}
\usepackage{array}

\usepackage[colorlinks=true,linktocpage=true,linkcolor=blue,citecolor=blue,hyperfootnotes=true]{hyperref}
\usepackage[normalem]{ulem}  
\usepackage{subfigure}

\begin{document}
	
%
\title{Nonlocal quark model description of a composite Higgs particle}
\author{Aliaksei Kachanovich}
\affiliation{Institute of Theoretical Physics, University of Wroclaw, PL-50204 Wroclaw, Poland}

\author{David Blaschke}
	\affiliation{Institute of Theoretical Physics, University of Wroclaw, PL-50204 Wroclaw, Poland}
	\affiliation{Bogoliubov Laboratory of Theoretical Physics, JINR Dubna, RU-141980 Dubna, Russia}
    \affiliation{National Research Nuclear University (MEPhI), RU-115409 Moscow, Russia}
	\date{\today}
\begin{abstract}
We propose a description of the Higgs boson as top-antitop quark bound state within a nonlocal  relativistic quark model of Nambu -- Jona-Lasinio type.  
In contrast to models with local four-fermion interaction, in the nonlocal generalization the mass of the scalar bound state can be lighter than the sum of its constituents. 
A simultaneous description of the experimentally determined values for both, the top quark mass and the scalar Higgs boson mass, is achieved by adjusting the interaction range and the value of the coupling constant. 
	\end{abstract}
\maketitle
	\section{Introduction}
	\label{intro}
The Higgs boson is a key ingredient of the Standard Model (SM), because it provides a description of 
spontaneous electro-weak $SU(2)_{L}\times U(1)_{Y}$ symmetry breaking \cite{Englert:1964et,Higgs:1964pj,Guralnik:1964eu}, for a review see \cite{Weinberg:1996kr}. 
Almost half a century of experimental quest for this last undiscovered piece of the SM  ended in 
2012 when ATLAS and CMS experiments at CERN announced the discovery of new particle, which matches requirements for the Higgs boson \cite{Aad:2012tfa,Chatrchyan:2012xdj}. 
The measured decay rates, and in particular two photon branching ratio, suggest that it is a short lived state of spin 0 and mass $125$~GeV.
Despite this obvious success, the theoretical interpretation of this scalar field as a fundamental object leads to a number of conceptual problems:
naturalness problem \cite{ArkaniHamed:2012kq}, problem of triviality \cite{Lane:2002wv} and the hierarchy problem.

An alternative to the elementary Higgs was proposed independently by Weinberg \cite{Weinberg:1975gm} and Susskind \cite{Susskind:1978ms}. 
They proposed new fermions which were coupled by a new strong force into a scalar particle. 
The interaction and the fermions were named technicolor and techniquarks, respectively. 
The theory predicts masses of the "ordinary" quarks and leptons, which match phenomenological data after complicated manipulations \cite{Dimopoulos:1979es,Eichten:1979ah}. 
Even after introducing these extensions technicolor had problems with flavour changing neutral currents (FCNC) which have been solved only for the first two families by introducing the walking coupling constant  \cite{Appelquist:1986tr,Holdom:1981rm,Yamawaki:1985zg}. 
With technicolour there occurred also a  problem with precision electroweak measurements \cite{Altarelli:1990zd,Golden:1990ig,Holdom:1990tc,Peskin:1990zt}.   
For a recent description of the $125$~GeV Higgs boson as a light technicolor scalar within a walking technicolor approach see, e.g., \cite{Foadi:2012bb}.

A different approach is the minimal supersymmetric standard model (MSSM) \cite{Dimopoulos:1981zb}. Supersymmetry could be a solution of the naturalness problem, because radiative corrections of fermions are divergent only logarithmically. 
Through supersymmetry transformations scalar loop corrections follow the same pattern. 
However, a general problem with supersymmetric models is that they predict superpartner particles for all known bosons and fermions with the same mass spectrum, distinguished from each other only by their spin. 
This has not  been experimentally observed yet. 
This problem can be avoided by a very high ($\gg 1$ TeV) supersymmetry violation scale. 
There are various supersymmetric models which predict measurable phenomena (see, e.g., \cite{Athron:2016usd,An:2012vp}).

There are also many exotic approaches to Higgs boson ranging from interpretations as a topological object to models which avoid the naturalness problem by introducing extra dimensions (for a review see, e.g., \cite{Csaki:2015hcd}).   

The first attempts to relate the symmetry breakdown of the standard model with a dynamical mechanism 
that generates the quark masses and describes the Higgs as a scalar composite particle have been made by Terazawa et al.~\cite{Terazawa:1976xx} in the attempt to unify particles and forces on the subquark level \cite{Terazawa:1979pj} by employing a model of the Nambu--Jona-Lasinio type 
\cite{Nambu:1961tp,Nambu:1961fr}.
Focussing on the dynamical chiral symmetry breaking in the top quark sector  Nambu \cite{Nambu:1989jt}, Miransky et al. \cite{Miransky:1988xi,Miransky:1989ds}
and Bardeen, Hill and Lindner \cite{Bardeen:1989ds} 
developed further the idea to
use the local Nambu -- Jona-Lasinio (NJL) model for a description of the mechanism for generating the bare mass $m_t$ of the top quark and the Higgs boson as a scalar $\bar{t} t$ bound state with a mass of $2m_t$ before renormalization-group corrections
(see, e.g., Ref.~\cite{Klevansky:1992qe} for a review on this model for dynamical chiral symmetry breaking and mesonic bound state generation in the light quark sector of low-energy QCD).
They also described the quark loop corrections to the $W^\pm$ and $Z^0$ boson masses within the renormalization-group setting of the  standard model and gave predictions for both, the Higgs boson and the top quark masses. 
At the time when Ref.~\cite{Bardeen:1989ds} was written, the mass of Higgs boson and the mass of the top quark were unknown. 
Nowadays it is plain that the application of the local NJL model is not consistent with the data, basically 
because of the scalar meson bare mass formula which yields $2m_t$ for the composite Higgs mass. 

In this paper we propose to revisit the idea of a composite Higgs boson within a nonlocal generalization of the Nambu model, which describes a scalar quark-antiquark bound state with a mass that is below the sum of the constituent quark masses. 
This has been demonstrated for the light quark sector in ref.~\cite{Schmidt:1994di}. 
In this paper we shall apply the nonlocal NJL model for the first time to the problem of spontaneous top quark mass generation and composite Higgs boson as a scalar $\bar{t} t$ bound state.
We shall demonstrate within a single-flavor model that the physical top and Higgs masses as they are known now from experiment can be described simultaneously.
The generalization of such a model to the flavor doublet structure of the standard model is the straightforward along the lines of \cite{Bardeen:1989ds}. This next step is deferred to future work.
	
\section{Nonlocal Nambu Quark Model for Scalar Bound State}
\label{Non-local Nambu Quark Model for Scalar Bound State}
	
In order to introduce the nonlocal effective theory we consider the ansatz of a local current-current vertex, but with nonlocal particle currents. 
Early versions of the nonlocal generalization of the NJL model have been given, e.g., in 
\cite{Schmidt:1994di,Bowler:1994ir,Anikin:1995cf}, see also \cite{Dorokhov:2003kf}.
We will follow here the introduction of the covariant separable model in  \cite{GomezDumm:2006vz},
but specialize it to the instantaneous case as in \cite{Schmidt:1994di}.
The effective action for non-local Nambu model has the form
\begin{equation}
S = \int d^{4} x \left[ \bar{q} (x) (-i \partial \!\!\!/) q (x) - \frac{G}{2} J(x)J(x) \right]~,
\end{equation}
 where as the main difference between to the local NJL models for the current $J(x)$ a nonlocal generalization is introduced in the form \cite{GomezDumm:2006vz}
\begin{equation}
J(x) = \int d^{4}z \ g(z)\  \bar{q}(x+\frac{z}{2})q(x-\frac{z}{2}),
\end{equation}	
with $g(z)$ being a formfactor which is responsible for the spatial nonlocality of the current.
Local models are a limiting cases of this form for $g(z)=\delta^{(4)}(z)$. 
We will choose specific ans\"atze for this form factor when performing the numerical solutions in 
Sect.~\ref{sec:results}. 

The quark mass is described by a gap equation and the mass of mesonic bound states is obtained 
from the Bethe-Salpeter equation in the corresponding interaction channel. 
For the derivation of these equations we use standard procedures \cite{Klevansky:1992qe} (for the nonlocal generalization see, e.g., \cite{Schmidt:1994di}).

Using Hubbard--Stratonovich bosonisation \cite{Hubbard:1959ub,Stratonovich:1957}, 
we obtain the action in the form
\begin{eqnarray}
S/V^{(4)} &=& - \mathrm{tr} \left\{ \ln \left[ (\gamma^{\mu} k_{\mu}) +   g(k) \sigma  \right]\right\}
 +\frac{\sigma^2}{4 G} .
\end{eqnarray}
where $\mathrm{tr}\{\dots\}=N_c \int \frac{d^{4} k}{(2\pi)^4} \mathrm{tr}_D\{\dots\}$ strands for the trace 
in color, Dirac and 4-momentum spaces, respectively.
Next we expand the effective action in powers of the fluctuation $\delta\sigma$ of the scalar meson field 
$\sigma$ around its vacuum expectation value $v$, 
\begin{equation}
\sigma = v + \delta \sigma,
\end{equation}
and obtain up to quadratic order
\begin{eqnarray}
\label{action}
S/V^{(4)} &=& \frac{v^{2}}{4G} + \frac{v \delta\sigma}{2 G} + \frac{(\delta \sigma)^{2} }{4G} 
- \mathrm{tr} \left[\ln G^{-1}_{MF}(k,k_0)\right] 
- \mathrm{tr} \left[G_{MF}(k,k_0)  g(k) \right] \delta\sigma 
\nonumber \\ 
&-& \frac{1}{2} \mathrm{tr} \left[g(k) G_{MF}(k,k_0) \delta\sigma \;g(k) G_{MF}(k,k_0) \delta\sigma\right]~,
\end{eqnarray} 
where the inverse mean field propagator
is
\begin{equation}
G_{MF}^{-1}(k,k_0) = \gamma^{\mu} k_{\mu} - m(k) ,
\end{equation}
with the dynamical quark mass
\begin{equation}
\label{mass}
{m}({k}) = g({k}){v}~.
\end{equation}
In order to assure stationarity of the action (\ref{action}) with respect to small fluctuations about $v$, the contribution linear in $\delta\sigma$ has to vanish.
This condition defines the meanfield gap equation which after performing Dirac trace, color summation 
and $k_0$ integration \cite{Klevansky:1992qe} takes the form
\begin{equation}
\label{gap}
v = 2 G N_{c} \int \frac{d^{3} k}{(2 \pi)^{3}} g^{2}(k) \frac{v}{E(k)},
\end{equation}
where $N_{c}$ is the number of colors, $g(k)$ is the form factor, which depends of a regularization method, and $E(k)$ is the relativistic dispersion relation
\begin{equation}\label{Relativistic energy}
E(k) = \sqrt{k^{2}+m^2(k)}.
\end{equation} 

The scalar quark-antiquark bound state which we shall denote as Higgs boson is described by the 
Bethe-Salpeter equation for the nonlocal theory. 
The inverse propagator $G_{\rm H}^{-1}(q,q_0)$ of the Higgs boson is defined by the terms in the action (\ref{action}) which are quadratic in the sigma field fluctuations 
\begin{eqnarray}
G_{\rm H}^{-1}(q,q_0)=\frac{1}{2G} - \Pi_{\rm H}(q,q_0) ~,
\end{eqnarray}
where the polarisation function is defined as
\begin{equation}
\label{polarization}
\Pi_{\rm H}(q,q_0) =  \mathrm{tr}\left[g(k) \; G_{MF} (k,k_0) \; g(k + q) \; G_{MF} (k + q,k_0+q_0) \right].
\end{equation}
The scalar meson mass is obtained by the mass pole condition defined with the polarization function for
a meson at rest ($q=0$) as
\begin{equation}
1-2G \Pi_{\rm H}(0,m_{\rm H}) = 0.
\end{equation}
After evaluating the traces implied in the definition (\ref{polarization}) and introducing the notation for averages of a momentum-dependent quantity $A({p})$ (for details, see \cite{Schmidt:1994di}),
\begin{equation}
\langle \langle A \rangle \rangle = \left[ \int dp p^{2} \frac{g^{2}(p)}{E(p)} \frac{A({p})}{E^{2}(p) - m_{\rm H}^{2}/4} \right] 
\left[ \int dp p^{2}  \frac{g^{2}(p)}{E(p)} \frac{1}{E^{2}(p) - m_{\rm H}^{2}/4}  \right]^{-1}~,
\end{equation}
we obtain the Higgs boson mass formula
\begin{eqnarray}
\label{mass-formula}
m_{\rm H}^{2} &=& 4 m^{2}(0) 
- 4 \langle \langle m^{2} (0) - m^{2} (p) \rangle \rangle .
\end{eqnarray}
This Higgs boson mass formula (\ref{mass-formula}) is the key result of this paper.
It explains how for a nonlocal quark model the mass of the scalar bound state can be lighter than the sum of the masses of its quark constituents due to their momentum dependence.

When the quark mass function drops with increasing momentum as in the cases of the example form factors of our nonlocal quark model, the second term in Eq.~(\ref{mass-formula}) leads to a reduction of the scalar meson mass and makes it a true bound state with a finite binding energy. 
This fact allows the simultaneous description of the Higgs boson as a scalar bound state of a top-antitop quark pair with their physical masses, for which holds $m_{\rm H}=0.718~m_t$.
This means that the Higgs boson as a composite of two top quarks is a strongly bound state!  
In the following section we perform the corresponding numerical solutions for this model for three examples of formfactor functions defining the nonlocal model and demonstrate that indeed the wanted solutions can be found. 

\section{Results}
\label{sec:results}

In the chiral limit, quarks remain massless until a critical value for the dimensionless coupling 
$G \Lambda^{2}$ is reached.
This critical value $G_c \Lambda^{2}$ depends on the choice of the form factor.
In our model calculations we will employ two types of form factors, namely
the Gaussian and generalized Lorentzian type,
\begin{eqnarray}
g_{G}(k) &=& \exp \Big( - \frac{k}{\Lambda_{G}}\Big)^{2} \\
g_{L}(k) &=&  \frac{1}{ 1 + \big(\frac{k}{\Lambda_{L}}\big)^{2\alpha}},
\end{eqnarray}
where $\Lambda$ is the effective range of the interaction in momentum space and $\alpha$ is a parameter regulating the shape of the generalized Lorentzian formfactor. The latter facilitates the regularization of the otherwise divergent loop integrals for the original Lorentzian formfactor ($\alpha=1$).

\begin{figure}[!htb]
\includegraphics[width=\textwidth]{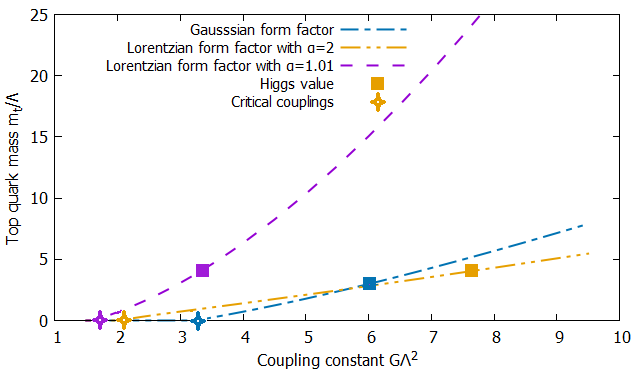}
\caption{Dimensionless mass of the top quark vs. dimensionless coupling constant $G\Lambda^{2}$. 
The square boxes indicate the values of the dimensionless coupling for which the ratio of Higgs mass to 
top quark mass assumes the physical value, see Fig.~\ref{fig:2}. }
\label{fig:1}   
\end{figure}

The solutions of the quark mass gap equation (\ref{gap}) are shown in Fig.~\ref{fig:1} for the top quark mass $m_t=m(k=0)=g(0)v=v$ in units of the interaction range $\Lambda$.
The values of the critical coupling are indicated by the fancy cross in Fig.~\ref{fig:1} and their values are given in the first column of Tab.~\ref{tab:1}. 

\begin{table}[!htbp]
\begin{tabular}{l|cccc}
\hline \hline
 & \hspace{5mm} $G_c \Lambda^{2}$ \hspace{5mm} & \hspace{5mm} $G \Lambda^{2}$ \hspace{5mm} & \hspace{5mm} $\Lambda$ \hspace{5mm} &\hspace{5mm}  $G $ \hspace{5mm} \\
 & & & [GeV]  & $ [10^{-3}$ GeV$^{-2}$] \\
 \hline
Lorentzian,  $\alpha = 1.01$ & 1.70 & 3.35&42.21&1.88\\
Lorentzian,  $\alpha = 2.00$ & 2.10 & 7.74&42.85&4.16\\
Gaussian & 3.28 & 6.01 & 58.10&1.79\\
\hline \hline
\end{tabular}
\caption{Dimensionless critical coupling constant $G_c\Lambda^2$ for the onset of spontaneous chiral symmetry breaking (first column), the value of the dimensionless coupling for which the physical ratio of Higgs to top quark mass is obtained (second column, see Fig.~\ref{fig:2}), the cut-off parameter $\Lambda$ which follows for this value and the physical top quark mass (third column) and corresponding coupling constant $G$ (last column). All values are given for three different form factors $g(x)$.}
\label{tab:1}
\end{table}

Only massive quarks can form a bound state. 
The maximal mass of a true bound state is given by the sum of the masses of its constituents which for a quark-antiquark bound state is twice the constituent mass.

\begin{figure}[!htb]
\includegraphics[width=\textwidth]{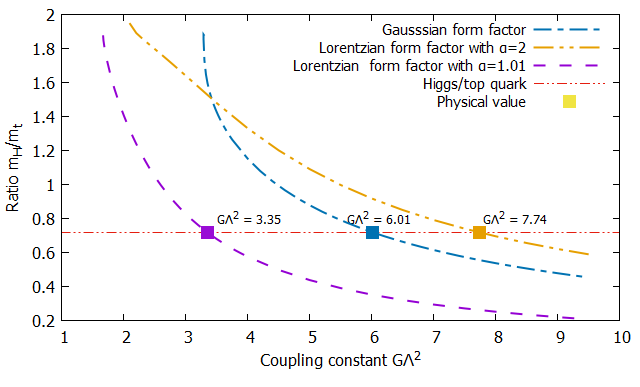}
\caption{Ratio of the Higgs boson and top quark masses in dependence of the dimensionless coupling 
$G\Lambda^2$ for three form factor models. The square boxes denote the values of the dimensionless coupling for which the mass ratio assumes the physical value, shown by the thin, red dash-double-dotted line.}
\label{fig:2}
\end{figure}

The ratio of the Higgs boson mass to the top quark mass is $m_{\rm H}/m_{t} = 0.718$. 
In Fig.~\ref{fig:2} this value is marked by a thin red dash-double dotted line. 
We show in this figure the possible values of the ratio $m_{\rm H}/m_{t}$ for different formfactors. 
For all formfactors one finds a value for $G\Lambda^2$ matching the experimental constraint. 
Using in addition the dependence of the dimensionless top quark mass $m_t/\Lambda$ on 
$G\Lambda^2$ as shown in Fig.~\ref{fig:1}, one can fix the cut-off parameter $\Lambda$ and the coupling constant $G$. 
The resulting values are presented in table \ref{tab:1}.

The values of coupling constant $G$ are about two orders of magnitude larger than the Fermi constant of the weak interaction $G_{F} = 1.664 \times 10^{-5} \mathrm{GeV^{-2}}$. This fact may be attributed to the nonlocality of the model as opposed to the local Fermi model. It would be interesting in an extension of this study to investigate other formfactors of the nonlocal NJL model approach. 

\section{Conclusions}
\label{Conclusions}
We have revisited the dynamical top quark mass generation mechanism that was studied by Bardeen, Hill and Lindner within the local NJL model now within a nonlocal generalization and obtained a composite Higgs boson as a scalar top-antitop quark bound state with a mass well below that of the local NJL mass formula for which $m_H=2 m_t$ before renormalization-group correction.
We isolated the effect of nonlocality in a generalized Higgs boson mass formula that allowed us to describe simultaneously the now experimentally known masses of the Higgs boson and the top quark, with a ratio
$m_{\rm H}=0.718~m_t$, by properly choosing the two free parameters of the single-flavor nonlocal NJL model.
In this way we could overcome an obstacle for the previous local NJL model setup and can proceed with a nonlocal generalization of \cite{Bardeen:1989ds} in subsequent work. 
There are other recent revivals of the top quark condensation approach to quark mass generation which deserve attention (see \cite{Zubkov:2014nka} predicting $m_{\rm H}=m_t/\sqrt{2}$) but which go beyond the scope of the present letter.

\section{Acknowledgements}
This work was supported by the Polish National Science Center (NCN) under grant No. UMO-2011/02/A/ST2/00306. The work of D.B. was supported in part by the MEPhI Academic Excellence Project under Contract No. 02.a03.21.0005.

\end{document}